\documentclass[]{JHEP3}

\usepackage{epsfig}
\usepackage{amsmath,amssymb}
\usepackage{citesort}

\allowdisplaybreaks[1]

\newcommand{\nn}{\nonumber}
\newcommand{\qq}{q\bar{q}}
\newcommand{\go}{\tilde{g}}
\newcommand{\gogo}{\go\go}
\newcommand{\sq}{\tilde{q}}

\newcommand{\simgt}
{\hbox{ \raise3pt\hbox to 0pt{$>$}\raise-3pt\hbox{$\sim$} }}
\newcommand{\simlt}
{\hbox{ \raise3pt\hbox to 0pt{$<$}\raise-3pt\hbox{$\sim$} }}

\title{Bound-state effects on gluino-pair production at hadron
       colliders}

\author{Kaoru Hagiwara\\
        KEK Theory Center and Sokendai, Tsukuba 305-0801, Japan}
\author{Hiroshi Yokoya\\
        Theory Unit, Physics Department, CERN, CH-1211 Geneva,
	Switzerland\\
        E-mail: \email{hiroshi.yokoya@cern.ch}}

\abstract{
We study bound-state effects on the pair production of gluinos at hadron
 colliders, in a context of the minimal supersymmetric extension of the
 standard model.
Due to the expected large mass and the octet color-charge of gluinos,
the bound-state effects can be substantial at the LHC.
We find significant deformation of the invariant-mass distributions of a
gluino-pair near the mass threshold, as well as an additional correction
to the total cross-section.
Both the invariant-mass distribution and the correction to the total
cross section depend crucially on the decay width of the gluino.
}

\keywords{Hadronic Colliders, Supersymmetric Standard Model, QCD}
\preprint{KEK-TH-1330\\ CERN-PH-TH/2009-169 \\ September 21, 2009}

\begin{document}

\section{Introduction}

At the CERN Large Hadron Collider (LHC), discovery of new heavy
 particles at TeV scale is expected as an evidence of physics beyond the
 Standard Model (SM).
One of the well-investigated models beyond the SM is the minimal
 supersymmetric extension of the Standard Model (MSSM).
At hadron colliders, colored particles can be produced copiously.
In the MSSM, the colored particles are superpartners of quarks and
 gluons, which are called squarks and gluinos, respectively.
The LHC has high discovery potential for those particles, where masses
 up to 2 TeV can be explored~\cite{tdr,Ball:2007zza}.
Current limits on their mass and production cross-sections
 are set by measurements at the Fermilab Tevatron
 collider~\cite{:2007ww,Aaltonen:2008rv}
(See also Ref.~\cite{Amsler:2008zzb}).\\

In the MSSM, due to the $R$-parity conservation, superpartner particles
 or sparticles, are produced in pair.
The total and differential cross-sections at hadron colliders have been
 known for long time in the leading-order (LO)~\cite{Harrison:1982yi,%
Dawson:1983fw,Haber:1984rc}, as well as
 in the next-to-leading oder (NLO) including SUSY-QCD corrections~\cite{%
 Beenakker:1996ch,Beenakker:1996ed,Beenakker:1997ut}.
The NLO corrections are found to positive and large.
The origin of the large corrections comes from the threshold logs due to
 the emission of the soft/collinear gluons, and also a Coulomb
 singularity due to the exchange of Coulomb gluons between the
 final-state particles.
Both terms become significant at the partonic threshold region,
 where they should be considered to all-order.
Recently, the all-order summation of the threshold logarithmic
 corrections has been performed for the sparticle-pair production
 processes at hadron colliders in
 Refs.~\cite{Kulesza:2008jb,Langenfeld:2009eg,Kulesza:2009kq}.\\

The aim of this paper is to examine the effects of Coulomb corrections
 to all-orders in the gluino-pair production process.
Similar studies for gluino-pair and also squark-pair productions can
 be found in Ref.~\cite{Kulesza:2009kq}, where the Coulomb corrections
 are taken into account to all-orders by using the Sommerfeld
 factor~\cite{Fadin:1990wx,Catani:1996dj}.
However, to our knowledge, there has been no study including the
 bound-states effect which have been found significant for the top-quark
 pair production at the
 LHC~\cite{Fadin:1990wx,Hagiwara:2008df,Kiyo:2008bv}.~\footnote{
We are informed that the bound-state effect on the squark pair
 production is examined in Ref.~\cite{Bigi:1991mi}.}
The bound-states of a pair of gluinos are called
 {\it gluinonium}~\cite{Nanopoulos:1983ys,Kuhn:1983sc,Keung:1983wz,%
 Goldman:1984mj,Kartvelishvili:1988tm,Hagiwara:1990sq,Chikovani:1996bk,%
 Khoze:2001xm,Cheung:2004ad}.
When the Coulomb force is attractive, it causes not only the corrections
 to the cross-sections above the mass threshold~\cite{Kulesza:2009kq},
 but also formation of the gluinoniums below the threshold.
The bound-state formations can be taken into account by summing the
 Coulomb terms to all-orders or non-perturbatively.
The Green's function formalism~\cite{Fadin:1987wz,Fadin:1988fn} has been
 developed for this purpose, which incorporates the finite-width effects
 of the constituent particles.

The paper follows the recent studies for the bound-state effects in the
 top-quark pair production at hadron
 colliders~\cite{Hagiwara:2008df,Kiyo:2008bv}.
The invariant-mass distribution of the top-quark pair is distorted
 significantly near the threshold
 region~\cite{Hagiwara:2008df,Kiyo:2008bv}, and we may expect similar
 effects for the gluino-pair.\\

The paper is organized as follows: In Sec.~\ref{sec:dec} and
 Sec.~\ref{sec:pro}, we briefly review the basic properties of
 the gluino decay-width and the gluino-pair production cross-section,
 respectively.
Then in Sec.~\ref{sec:bind}, we discuss binding corrections to the
 gluino-pair production.
 In Sec.~\ref{sec:isr}, we discuss effects of the initial-state
 radiation and the hard-vertex correction.
Finally, we summarize our findings in Sec.~\ref{sec:sum}.

\section{Gluino decay-width}\label{sec:dec}

In this section and the next section, we briefly review the basic
 properties of the gluino that are relevant to our studying; the
 decay-width and the pair-production cross-section at hadron colliders,
 respectively.
For more details, see e.g.\ Ref.~\cite{Drees:2004jm}.\\

The effects of gluinonium formation at hadron colliders depend strongly
 on the gluino decay width.
In particular, the following regions should have qualitatively different
 signals:
\begin{subequations}
\begin{align}
 &A: \quad \Gamma_{\go}\gtrsim |E_{B}| \label{reg:a}\\
 &B: \quad |E_{B}| > \Gamma_{\go}\gg \Gamma_{gg}  \label{reg:b}\\
 &C: \quad |E_{B}| \gg \Gamma_{\go} > \Gamma_{gg} \label{reg:c}\\
 &D: \quad \Gamma_{\go}< \Gamma_{gg} \label{reg:d}
\end{align}\label{reg}
\end{subequations}
Here, $E_{B}<0$ is the binding-energy of gluinoniums, and $\Gamma_{gg}$
 denotes the partial decay-width of the gluinonium annihilation into
 gluons.
In the region $A$, the produced gluinos decay before the gluinonium
 formation, and the binding effects cannot be observed.
In the region $B$, the binding effects are expected to enhance the pair
 production cross-section below the threshold, especially at around the
 location of the ground state energy, as is expected for the
 top-quarks~\cite{Fadin:1990wx,Hagiwara:2008df,Kiyo:2008bv}.
In the region $C$, a few narrow gluinonium resonances can be produced,
 whose decay may still be dominated by the constituent gluino decays.
In the region $D$, the produced gluinonium will decay mainly into gluon
 jets, and hence will disappear without leaving a detectable trace in
 hadron collider environments.\\

\FIGURE[h]{
\epsfig{file=width.eps,width=0.45\textwidth}
\caption{
The gluino decay-width as a function of the gluino mass, calculated from
 $\go\to q\bar{q}^{(}{'}{}^{)}\tilde{W}$ and $\go\to q\bar{q}\tilde{B}$
 decay modes.
We consider decays into 5-flavor massless quarks only where the
corresponding squarks have a common mass $m_{\sq}=500$ GeV, 1 TeV or
 1.5 TeV, and the gaugino masses satisfy
 $m_{\go}:m_{\tilde{W}}:m_{\tilde{B}}=7:2:1$.
The dotted-lines show the magnitude of the binding energy of the ground
 $^{1}S_{0}({\bf 1})$ gluinonium ($-E_{B}[^{1}S_{0}]$), and the partial
 decay-width of the
 ground $^{1}S_{0}({\bf 1})$ gluinonium annihilation into gluons
 ($\Gamma_{gg}[^{1}S_{0}]$).\label{fig:width}}
}

In Fig.~\ref{fig:width}, we plot the decay width of the gluino as a
 function of the gluino mass.
The decay-width is calculated for $\go\to
 q\bar{q}^{(}{'}{}^{)}\tilde{W}$ and $\go\to q\bar{q}\tilde{B}$
 decays~\cite{Barnett:1987kn}, where winos ($\tilde{W}$) and a
 bino ($\tilde{B}$) are superpartners of SU(2)$_{\rm L}$ and U(1)$_{\rm
 Y}$ gauge bosons, respectively.
Solid, dashed and dot-dashed curves are for the squark mass
 $m_{\sq}=500$ GeV, 1 TeV and 1.5 TeV, respectively.
For simplicity, we assume that all the super-partners of 5 light-quarks
 with both chiralities have a common mass $m_{\sq}$, and the gaugino
 masses satisfy the relation
 $m_{\go}:m_{\tilde{W}}:m_{\tilde{B}}=7:2:1$, which are valid in several
 SUSY breaking scenarios.
Neither the contribution from the top squarks nor the gaugino-higgsino
 mixing are considered for brevity.

The upper dotted-line shows the magnitude of the binding energy of the
 ground $^{1}S_{0}(\bf 1)$ gluinonium, which is the most deeply-bounded
 color--singlet gluinonium, see section~\ref{sec:bind} in detail.
It is estimated by using a Coulombic potential $V(r)=-C_A\alpha_s/r$
 with the color-factor $C_A=3$ for the color--singlet gluinoniums.
Within this approximation, the binding-energy is calculated as
\begin{align}
 E_{B}[^{1}S_{0}({\bf
 1})]=-\frac{C_A^2}{4}m_{\go}\alpha_s^2(\mu_B)_{\overline{\rm MS}},
\end{align}
 where we take the ${\overline{\rm MS}}$ renormalization scale to
 satisfy $\mu_{B}=C_Am_{\go}\alpha_s(\mu_{B})_{\overline{\rm MS}}/4$,
 which corresponds to a half of the inverse Bohr radius of the
 $^{1}S_{0}({\bf 1})$ gluinonium.
It is known that this scale choice makes the QCD higher-order
 corrections to the binding energy small.
For the gluino-mass from 200 GeV to 2 TeV, the binding energy grows from
 $\sim 10$ GeV to $\sim 50$ GeV.

The lower dotted-line shows the partial decay-width of the ground
 $^1S_0({\bf 1})$ gluinonium annihilation into two gluons;
\begin{align}
 \Gamma_{gg}[^{1}S_{0}({\bf1})] =
 18\pi\alpha_s^2(\frac{m_{\go}}{2})_{\overline{\rm MS}}\cdot
 \frac{\left|\psi(0)\right|^2}{m_{\go}^2},\label{eq:ggg}
\end{align}
 with
 $\left|\psi(0)\right|^2=C_A^3m_{\go}^3\alpha_s^3(\mu_{B})_{\overline{\rm
 MS}}/8\pi$ which is also obtained by using the Coulombic potential
 approximation.
For the same gluino-mass region as above, it grows from several hundred
 MeM to one GeV. \\

The three curves for $\Gamma_{\go}$ show steep gradient at
 $m_{\go}\simeq m_{\sq}$.
Above the squark mass threshold, the gluinos decay into a squark and a
 quark and the width is proportional to $\alpha_sm_{\go}$.
Below the two-body decay threshold, the gluino decay width drops
 quickly down to the three-body decay width with the electroweak
 coupling.
The gluino width curves cross the binding-energy curve at around
 $m_{\go}\simeq 1.3m_{\sq}$, while they cross the gluinonium
 annihilation width just above $m_{\go}=m_{\sq}$.
Therefore, if the gluino is lighter than all the squarks, the region
 $D$~(\ref{reg:d}) would likely be realized, and binding effects cannot
 be observed at hadron colliders.
On the other hand for $m_{\go}>m_{\sq}$, depending on the mass ratio
 $m_{\go}/m_{\sq}$, regions $A$ to $C$~(\ref{reg}a-c) would be
 realized.

\section{Gluino-pair production cross-section}\label{sec:pro}

Now, we review the gluino-pair production at hadron colliders, focusing
 our attention on the threshold behavior and the color structure of the
 scattering amplitudes.\\

At the parton level, there are two leading subprocesses for the
 gluino-pair production;
\begin{subequations}
\begin{align}
 g(p_1,\lambda_1,a_1) + g(p_2,\lambda_2,a_2)
 &\to \go(p_3,\lambda_3,a_3) + \go(p_4,\lambda_4,a_4)\label{sub:gg},\\
 q(p_1,\lambda_1,i_1) + \bar{q}(p_2,\lambda_2,i_2)
 &\to \go(p_3,\lambda_3,a_3) + \go(p_4,\lambda_4,a_4)\label{sub:qq}.
\end{align}
\end{subequations}
The tree-level Feynman-diagrams are shown in Fig.~\ref{feyndiag}.
Here, $p_k$ and $\lambda_k$ are momenta and helicities of particles,
 $a_k$ and $i_k$ are color indices of gluons(gluinos) and quarks,
 respectively.
We normalize $\lambda_k$ to take $\pm 1$ both for fermions and
 gluons.
At the LHC, due to the steep rise of the gluon density in the small
 momentum-fraction region, gluon-fusion process (\ref{sub:gg})
 dominates the cross section for $m_{\go}<1$
 TeV~\cite{Beenakker:1996ch}.
For $m_{\go}>1$ TeV, the $\qq$ annihilation process (\ref{sub:qq}) can
contribute significantly to the total cross-section.

\FIGURE[t]{\epsfig{file=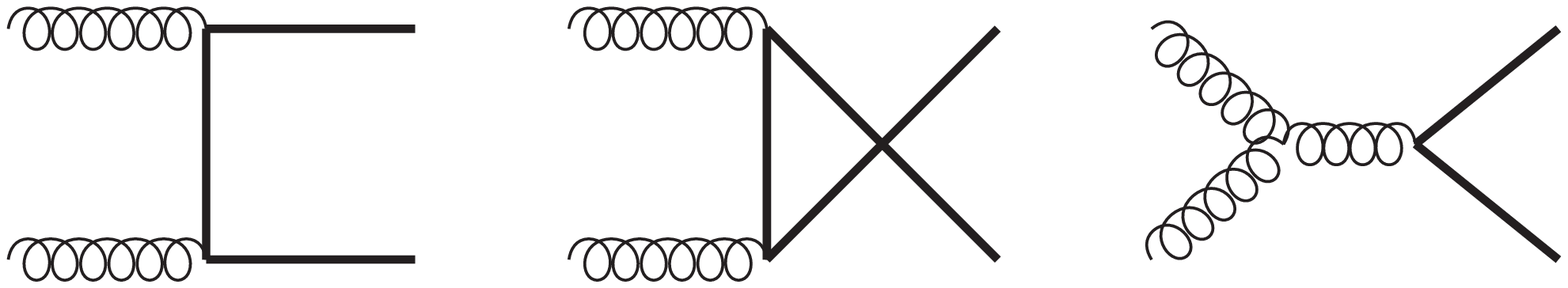,width=0.45\textwidth}\hspace{20pt}
\epsfig{file=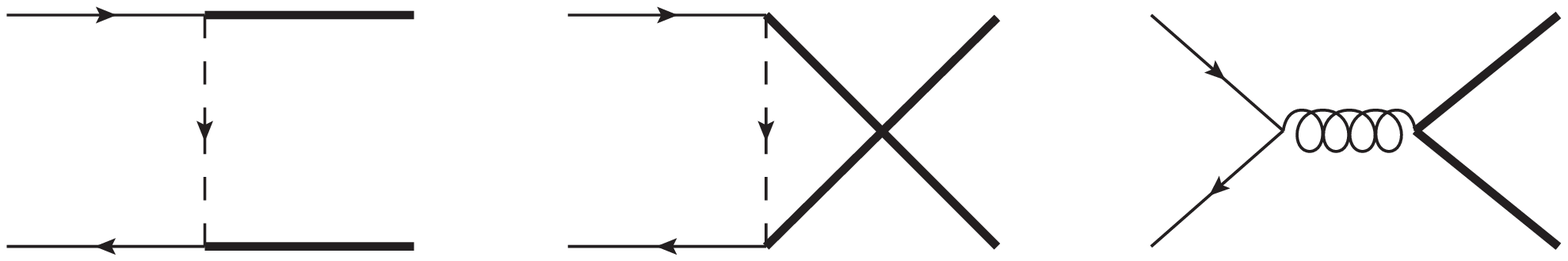,width=0.45\textwidth}
 \caption{Feynman diagrams for $gg\to\gogo$ process and $\qq\to\gogo$
 process at the tree-level.
 Dashed-lines represent squarks, $\tilde{q}_R$ or
 $\tilde{q}_L$.\label{feyndiag}}}

\subsection{$gg\to\gogo$ process}

First, we review the gluon-fusion process.
We present helicity amplitudes for this process by decomposing the
 color factors into symmetric and antisymmetric parts:
\begin{align}
 {\cal M}_{gg\to\gogo}(p_i;\lambda_i;a_i) = g_s^2\Big[
 \frac{1}{2}\left\{F^{a_3},F^{a_4}\right\}_{a_1a_2}
 M_{gg}(p_i;\lambda_i) +
 \frac{1}{2}\left[F^{a_3},F^{a_4}\right]_{a_1a_2}
 N_{gg}(p_i;\lambda_i)\Big].\label{eq:megogo}
\end{align}
Here $g_s=\sqrt{4\pi\alpha_s}$ is the QCD coupling constant.
$F^{a}_{bc}=-if^{abc}$ is the color matrix in the adjoint
 representation.
The amplitudes can be obtained by replacing only the color-factors from
 those of the heavy-quark pair production, $gg\to Q\bar{Q}$.
The color--symmetric part and antisymmetric part of the amplitudes are
 summarized as follows;
\begin{subequations}
\begin{align}
 &M_{gg}(p_k;\lambda,\lambda,\bar{\lambda},\bar{\lambda}) =
 \frac{4m_{\go}}{\sqrt{\hat{s}}}
 \frac{\lambda+\bar{\lambda}\beta}{1-\beta^2\cos^2\theta},\\
 &M_{gg}(p_k;\lambda,\lambda,\bar{\lambda},-\bar{\lambda}) = 0,\\
 &M_{gg}(p_k;\lambda,-\lambda,\bar{\lambda},\bar{\lambda}) =
 -\bar{\lambda} \frac{4m_{\go}}{\sqrt{\hat{s}}}
 \frac{\beta\sin^2\theta}{1-\beta^2\cos^2\theta}
 ,\\
 &M_{gg}(p_k;\lambda,-\lambda,\bar{\lambda},-\bar{\lambda}) =
 -2\frac{\beta\sin\theta(\lambda\bar{\lambda}+\cos\theta)}
 {1-\beta^2\cos^2\theta},
\end{align}
\end{subequations}
for the symmetric part, and
\begin{align}
 &N_{gg}(p_k;\lambda_k) = \beta\cos\theta\cdot
 M_{gg}(p_k;\lambda_k).\label{eq:rel}
\end{align}
 for the anti-symmetric part.
Here, $\theta$ is the angle between the 3-momenta $\vec{p}_1$ and
 $\vec{p}_3$, and $\beta=\sqrt{1-4m^2_{\go}/\hat{s}}$ is the velocity of
 the gluino in the partonic center-of-mass frame with
 $\hat{s}=(p_1+p_2)^2$.\\

The product of the two color--octet states transforms as ${\bf
 8}\otimes{\bf 8}={\bf 1}\oplus{\bf 8_S}\oplus{\bf 8_A}\oplus{\bf
 10}\oplus\overline{\bf 10}\oplus{\bf 27}$.
In the $gg\to\gogo$ amplitudes, the color basis for these states can be
 taken as~\cite{Kulesza:2008jb};
\begin{subequations}
\begin{align}
 {\bf 1}:\quad &c_1 = \frac{1}{8}\delta^{a_1a_2}\delta^{a_3a_4},\\
 {\bf 8_S}:\quad &c_2 = \frac{3}{5}d^{a_1a_2b}d^{a_3a_4b},\\
 {\bf 8_A}:\quad &c_3 = \frac{1}{3}f^{a_1a_2b}f^{a_3a_4b},\\
 {\bf 10\oplus\overline{10}}:\quad &c_4 = \frac{1}{2}\left(
 \delta^{a_1a_3}\delta^{a_2a_4}-\delta^{a_1a_4}\delta^{a_2a_3}\right)
 - \frac{1}{3}f^{a_1a_2b}f^{a_3a_4b},\\
 {\bf 27}:\quad &c_5 = \frac{1}{2}\left(
 \delta^{a_1a_3}\delta^{a_2a_4}+\delta^{a_1a_4}\delta^{a_2a_3}\right)
 - \frac{1}{8}\delta^{a_1a_2}\delta^{a_3a_4}
 - \frac{3}{5}d^{a_1a_2b}d^{a_3a_4b},
\end{align}\label{eq:basis}
\end{subequations}
 where the bases $c_1$ to $c_5$ are for ${\bf 1}$, ${\bf 8_S}$,
 ${\bf 8_A}$, ${\bf 10}\oplus\overline{\bf 10}$, ${\bf 27}$,
 respectively.
The color factors in Eq.~(\ref{eq:megogo}) are then expressed as
\begin{align}
 &\frac{1}{2}\left\{F^{a_1},F^{a_2}\right\}_{a_3a_4}
 = \frac{1}{2}
 \left(f^{a_1a_3b}f^{a_2a_4b}+f^{a_1a_4b}f^{a_2a_3b}\right)
 = 3c_1+\frac{3}{2}c_2-c_5,\\
 &\frac{1}{2}\left[F^{a_1},F^{a_2}\right]_{a_3a_4}
 = \frac{1}{2}
 \left(f^{a_1a_3b}f^{a_2a_4b}-f^{a_1a_4b}f^{a_2a_3b}\right)
 = \frac{3}{2}c_3.
\end{align}
The $c_4$ term for ${\bf 10}\oplus\overline{\bf 10}$ state does not
 appear at the tree-level.
Near the threshold, only the color-symmetric part of the amplitude
 $M_{gg}$ survives, and the anti-symmetric part $N_{gg}$ is suppressed
 by $\beta$.
Thus, production cross-section for the color-${\bf 8_A}$ state is
 suppressed by ${\cal O}(\beta^2)$ near the threshold.
All the other color--states have the same production amplitude at the
 tree-level, with the normalization constant
 $(|3c_1|^2,|3/2c_2|^2,|c_5|^2)=(9,18,27)$.
Therefore, 1/6 of gluino-pair produced near the threshold is
 in the color--singlet, while 1/3, 1/2 are in the ${\bf 8_S}$, ${\bf
 27}$ color--states, respectively.

\subsection{$\qq\to\gogo$ process}

Next, we examine the $\qq$ annihilation process.
In contrast to the $\qq\to Q\bar{Q}$ process where only the color--octet
 state are produced, the color--singlet $\gogo$ state can also be
 produced due to the squark-exchange diagrams; see
 Fig.~\ref{feyndiag}. \\

Helicity amplitudes for this process are written as;
\begin{align}
 {\cal M}_{\qq\to\gogo}(p_k;\lambda_k;i_k,a_k)
 = g_s^2\Big[\frac{1}{2}\left\{T^{a_3},T^{a_4}\right\}_{i_1i_2}
 M_{\qq}(p_k;\lambda_k) + \frac{1}{2}
 \left[T^{a_3},T^{a_4}\right]_{i_1i_2}
 N_{\qq}(p_k;\lambda_k)\Big].\label{eq:ampqq}
\end{align}
The color--symmetric amplitudes are
\begin{subequations}
\begin{align}
 &M_{\qq}(p_k;\lambda,-\lambda,\bar{\lambda},\bar{\lambda}) =
 \bar{\lambda}\frac{4m_{\go}}{\sqrt{\hat{s}}}
 \frac{\beta\cos\theta\sin\theta}
 {A_{\lambda}^2-\beta^2\cos^2\theta},\\
 &M_{\qq}(p_k;\lambda,-\lambda,\bar{\lambda},-\bar{\lambda}) =
 -2\beta(1+\lambda\bar{\lambda}\cos\theta)
 \frac{A_{\lambda}-\lambda\bar{\lambda}\cos\theta}
 {A_{\lambda}^2-\beta^2\cos^2\theta},
\end{align}
\end{subequations}
 and the anti-symmetric ones are
\begin{subequations}
\begin{align}
 &N_{\qq}(p_k;\lambda,-\lambda,\bar{\lambda},\bar{\lambda}) =
 \bar{\lambda}\frac{4m_{\go}}{\sqrt{\hat{s}}}\sin\theta
 \left[1-\frac{A_{\lambda}}{A_{\lambda}^2-\beta^2\cos^2\theta}\right],\\
 &N_{\qq}(p_k;\lambda,-\lambda,\bar{\lambda},-\bar{\lambda}) =
 2\left(\lambda\bar{\lambda}+\cos\theta\right)
 \left[1-\frac{A_{\lambda}-\lambda\bar{\lambda}\beta^2\cos\theta}
 {A_{\lambda}^2-\beta^2\cos^2\theta}\right].
\end{align}\label{eq:nqq}
\end{subequations}
Here, we define $A_{\lambda}=1-2(m^2_{\go}-m^2_{\sq_{\lambda}})/\hat{s}$
 where $\sq_{+}=\sq_{R}$ and $\sq_{-}=\sq_{L}$.
Due to the chirality conservation, amplitudes for $\lambda_1=\lambda_2$
 vanish and only the squark with the chirality of the incoming quark
 $\lambda=\lambda_{1}$ contributes.
In the threshold limit $\beta\to 0$, the color-symmetric amplitudes
 behave as $M_{\qq}\sim {\cal O}(\beta)$ and the anti-symmetric ones as
 $N_{\qq} \propto
 (m^2_{\go}-m^2_{\sq_{\lambda}})/(m^2_{\go}+m^2_{\sq_{\lambda}})$.
Note that, the first term in the square bracket in Eqs.~(\ref{eq:nqq})
 comes from the $s$-channel gluon exchange diagram, while the other term
 comes from squark-exchange diagrams.
When $A_{\lambda}= 1$ i.e.\ when the gluino mass and the squark mass are
 degenerate, the $s$-channel diagram and squark-exchange diagrams
 interfere destructively~\cite{Beenakker:1996ch} to make the full
 amplitudes (\ref{eq:ampqq}) color-symmetric.

The color states of the produced gluino pair are found from
\begin{subequations}
\begin{align}
 &\frac{1}{2}\{T^{a},T^{b}\}_{ij}
 =\frac{1}{2N}\delta^{ab}\delta_{ij}+ \frac{1}{2}
 d^{abc}T^{c}_{ij},\label{eq:cfs} \\
 &\frac{1}{2}[T^{a},T^{b}]_{ij} =
 \frac{1}{2}if^{abc}T^{c}_{ij}.\label{eq:cfa}
\end{align}
\end{subequations}
The first term in the right-hand side in Eq.~(\ref{eq:cfs}) represents
 the color--singlet, and the others are color--octets either symmetric
 (\ref{eq:cfs}) or anti-symmetric (\ref{eq:cfa}).
Thus, in $\qq$ annihilation, production of the color--singlet
 state is suppressed by ${\cal O}(\beta^2)$ and the color--octet state
 ($\bf 8_A$) dominates near the threshold, unless the gluino mass and
 the squark mass are degenerate.

\section{Binding corrections}\label{sec:bind}

In this section, we describe the bound-state effects in the
 gluino-pair production.
The partonic cross-section including bound-state correction is expressed
 as;
\begin{align}
 \hat\sigma^{(c)}_{i}(\hat{s}) = \hat\sigma^{(c)}_{i,0}(\hat{s})\cdot
 \frac{{\rm Im}[G^{(c)}(\vec{0},E+i\Gamma_{\go})]}
 {{\rm Im}[G_0(\vec{0},E+i\Gamma_{\go})]}, \label{eq:coul}
\end{align}
 where $\hat\sigma^{(c)}_{i,0}$ is the tree-level cross-section.
Here, we specify the color--state of the gluino-pair denoted by $c$, and
$i$ represents the initial partons $i=gg$ or $\qq$.
The Green's function $G^{(c)}$ is obtained by solving the Sch\"odinger
 equation;
\begin{align}
 \left[\left(E + i\Gamma_{\go}\right)
 -\left\{-\frac{\nabla^2}{m_{\go}}+V^{(c)}(r)\right\}\right]
 G^{(c)}(\vec{x};E+i\Gamma_{\go}) = \delta^{3}(\vec{x}),\label{eq:scho}
\end{align}
 where $E$ is an energy of the gluino-pair measured from their mass
 threshold.
We define it as,
\begin{align}
 E = \left\{
 \begin{array}{cc}
  M-2m_{\go} & (M<2m_{\go})\\
  m_{\go}\,\beta^2 & (M\ge 2m_{\go})\label{eq:ene}
 \end{array}
 \right. ,
\end{align}
 where $M$ is the invariant-mass of the gluino-pair.
$V^{(c)}(r)$ is QCD potentials between the gluino-pair depending on
 their colors.
$G_0$ is obtained by taking $V^{(c)}=0$, which is simply the tree-level
Green's function of the gluinos with its finite width.

We use the NLO QCD potential~\cite{Kniehl:2004rk};
\begin{align}
 V^{(c)}(r) =
 C^{(c)}\frac{\alpha_{s}(\mu_{B})_{\overline{\rm MS}}}{r}
 \left[1+\frac{\alpha_s}{\pi}\left\{2\beta_0\left[\ln{(\mu_{B} r)
 + \gamma_{E}}\right]+a_{1}\right\}
 +{\cal O}\Big(\big(\frac{\alpha_s}{\pi}\big)^2\Big)\right]
\end{align}
 where the color-factor $C^{(c)}$ is known~\cite{Goldman:1984mj} to be
\begin{align}
 C^{(c)}=\{-C_A,-\frac{1}{2}C_A,-\frac{1}{2}C_A,0,1\},\label{eq:cfac}
\end{align}
 for $c=\{{\bf 1, 8_S, 8_A, 10\oplus \overline{10}, 27}\}$,
 respectively.
Thus, QCD potential for {\bf 1}, ${\bf 8_S}$ and ${\bf 8_A}$
 color--states is attractive, while that for {\bf 27} color--state is
 repulsive.
$\gamma_{E}=0.5772\dots$ denotes the Euler constant, and the other
 constants are;
\begin{align}
 &\beta_0=\frac{11}{3}C_{A}-\frac{2}{3}n_{q},\quad
 a_{1}=\frac{31}{9}C_{A}-\frac{10}{9}n_{q}.
\end{align}
We take the number of light-quark flavor to $n_q=5$, and the scale
 $\mu_{B}$ as a half of the inverse Bohr radius of the color--singlet
 gluinonium, $\mu_{B}=1/2r_{B}=C_{A}m_{\go}\alpha_s(\mu_{B})/4$.
For example for $m_{\go}=608$ GeV, we obtain $\mu_{B}\simeq 58$ GeV and
 $\alpha_s(\mu_{B})\simeq 0.127$.
With this scale choice, the higher-order corrections to the binding
 energy is small.
On the other hand, the magnitude of the Green's function is reduced by
 several tens percent in an entire range of $E$, when one includes the
 NLO correction.\\

For the attractive force, the gluino-pair forms a bound-state with
 a definite spin ($S$) and an orbital angular momenta
 ($L$)~\cite{Goldman:1984mj}.
Within the MSSM the gluino is a Majorana fermion, and the wave-function
 of the $\gogo$ state must be anti-symmetric under the exchange
 of the two gluinos.
This leads to the restriction on the spin and orbital angular
 momenta of the gluinonium, so that the color--symmetric pair in ({\bf
 1}, ${\bf 8_S}$, {\bf 27}) forms only even $L+S$ states, while the
 color--antisymmetric pair (${\bf 8_A}$) forms only odd
 $L+S$ states.
Possible bound-states for each color--state are summarized
 in Table~\ref{tab:gg}.
We also list the non-zero components in the tree-level amplitude for
 each process.
Blank states in $i=gg$ and $i=\qq$ rows are missing at the tree-level.
For example, due to the Yang's theorem, color-symmetric $J=1$ states are
 forbidden in gluon-fusion process.
Likewise, $S=0$ or $J=0$ states are forbidden in $i=\qq$, due to the
 chirality conservation.

For the repulsive force, the gluino-pair does not form a bound-state,
 but the Green's function formula is still useful for summing over
 Coulombic corrections.

\renewcommand{\arraystretch}{1.2}
\TABLE[b]{\begin{tabular}{|c|c|c|}
 \hline
 color & symmetric ({\small ${\bf 1}$, ${\bf 8_S}$, ${\bf 27}$}) &
 anti-symmetric ({\small ${\bf 8_A}$}) \\
 \hline\hline $\gogo$ &
 $^{1}S_{0}$, $^{3}P_{0,1,2}$, $^{1}D_{2}$, $\cdots$ &
 $^{3}S_{1}$, $^{1}P_{1}$, $^{3}D_{1,2,3}$, $\cdots$ \\
 \hline\hline $i=gg$ &
 $^{1}S_{0}$, $^{3}P_{0,\hphantom{1,}2}$, $^{1}D_{2}$,
 $\cdots$ & \hphantom{$^{3}S_{1}$,}
 $^{1}P_{1}$, $^{3}D_{1,\hphantom{2,}3}$, $\cdots$ \\
 \hline
 $i=\qq$ & \hphantom{$^{1}S_{0}$,}
 $^{3}P_{\hphantom{0,}1,2}$, \hphantom{$^{1}D_{2}$,}
 $\cdots$ & $^{3}S_{1}$, \hphantom{$^{1}P_{1}$,}
 $^{3}D_{1,2,3}$, $\cdots$ \\
 \hline
\end{tabular}
\caption{Summary of gluinonium states, and possible tree-level
 production from $i=gg$ and $i=\qq$.\label{tab:gg}}}

In our analysis, we include only the $S$-wave Green's function.
The other states are left free, since they are not affected
 significantly by the binding effects.
The extraction of the certain partial-wave amplitude is straightforward
 from the helicity amplitudes given in the previous section. \\

The $\gogo$ invariant-mass distribution in the hadronic collisions is
 given as~\cite{Hagiwara:2008df,Kiyo:2008bv};
\begin{align}
 \frac{d\sigma}{dM^2} = \sum_{i,c}\hat{\sigma}^{(c)}_{i}(M^2)
 \cdot K^{(c)}_{i}
\int_{\tau}^{1}\frac{dz}{z}F^{(c)}_{i}(z)
 \frac{d{\cal L}_{i}}{d\tau}\left(\frac{\tau}{z}\right),\label{eq:fac}
\end{align}
 where $\tau=M^2/s$.
The partonic luminosity is defined by
\begin{align}
 \frac{d{\cal L}_i}{d\tau}(\tau;\mu_F)=\sum_{\{a,b\}}\int dx_1\int dx_2
 f_a(x_1,\mu_F)f_b(x_2,\mu_F)\delta{(\tau-x_1x_2)},
\end{align}
 where the summation is over $\{a,b\}=\{g,g\}$ for $i=gg$, and
 $\{a,b\}=\{q,\bar{q}\},\{\bar{q},q\}$ with $q=u,d,s,c,b$ for $i=\qq$.
We use 5-flavor parton distribution functions, and the 5-flavor strong
 coupling constant even for $\mu>m_t$.

The convolution formula in Eq.~(\ref{eq:fac}) is based on an assumption
 of the factorization of the soft/collinear gluon emission and the
 Coulomb corrections\footnote{%
Recent paper by M.~Beneke, P.~Falgari and C.~Schwinn shows the
 factorization property of the Coulomb-gluon and the
 soft-gluon~\cite{Beneke:2009rj}.}.
Thus, the soft/collinear gluon emission is described by the ISR
 (initial-state radiation) functions $F^{(c)}_{i}$, and the Coulomb
 corrections are included in the partonic cross-section
 $\hat\sigma^{(c)}_{i}$, as in Eq.~(\ref{eq:coul}).
$K^{(c)}_{i}$ represents the hard-vertex correction, which is a
 process-dependent constant.

In this section, we neglect the ISR effect for simplicity, and set
 $F^{(c)}_{i}(z)=\delta{(1-z)}$ with $K^{(c)}_{i}=1$.
The effect is discussed in the next section.\\

\FIGURE[t]{\epsfig{file=mglgl.eps,width=0.48\textwidth}\quad
 \epsfig{file=mglgl_c.eps,width=0.48\textwidth}
 \caption{$\gogo$ invariant-mass distribution at the LHC, for
 $m_{\go}=608$ GeV and $\Gamma_{\go}=5.5$ GeV (SPS1a).
 Result in the Born level (dotted), ${\cal O}(\alpha_s)$ Coulomb
 correction (dashed), and the all-order Coulomb correction (solid) are
 plotted.
 Right figure is the same, but enlarged for the threshold
 region.
 All-order Coulomb correction for individual {\bf 1} and
 ${\bf 8_S}$ color--states in the gluon-fusion process are also
 plotted in dot-dashed lines.\label{fig:inv}}}

In the left figure in Fig.~\ref{fig:inv}, we show the invariant-mass
 distribution of the gluino-pair including bound-state corrections, for
 the LHC with $\sqrt{s}=14$ TeV.
The results are given for $m_{\go}=608$ GeV, {\it a la}
 SPS1a~\cite{Allanach:2002nj}, and for $\Gamma_{\go}=5.5$ GeV for
 $m_{\sq}=547$ GeV.
We use the CTEQ6L PDF parameterization~\cite{Pumplin:2002vw}, and set
 scales as $\mu_R=\mu_F=m_{\go}$.
Predictions in Born-level, including ${\cal O}(\alpha_s)$ Coulomb
 correction, and all-order Coulomb correction are plotted in dotted,
 dashed and solid lines, respectively.
Compared to the Born-level result, the ${\cal O}(\alpha_s)$ correction
 as well as the all-order Coulomb corrections bring large enhancement
 near the mass threshold ($M\simeq 2m_{\go}$).
Moreover, the distribution for the all-order correction indicates
 the resonance peaks below the threshold.
Due to the binding correction and the finite-width effect, the
 gluino-pair invariant-mass distribution arises rapidly below the mass
 threshold.
Thus the effective pair-production threshold is not at $2m_{\go}$ but at
 $M\simeq 2m_{\go}-30$ GeV.
The order of the deviation is about the binding-energy of the
 $^{1}S_{0}({\bf 1})$ state, thus ${\cal O}(m_{\go}\alpha_s^2)$.
On the other hand, the enhancement disappears in the high invariant-mass
 region.

The all-order Coulomb correction enhance the total cross-section by
 about 15\% from the cross-section in Born level, where a substantial
 portion from below the threshold region contributes.
We discuss this result in more detail below. \\

The right figure in Fig.~\ref{fig:inv} is the same, but enlarged on the
 threshold region.
For the all-order correction, individual contributions of the color
 {\bf 1}, ${\bf 8_S}$ states are also plotted.
But those for other color--states, ${\bf 8_A}$, ${\bf 27}$ which have no
 enhancement near the threshold are not shown.
$\qq$ annihilation process is negligible in this case.
In the all-order distribution, two resonance peaks can be found at
 $E\simeq-22$ GeV and $E\simeq-7$ GeV.
The first peak corresponds to the ground state of the color--singlet
 $^{1}S_{0}$ resonance, and the second peak consists of the second
 excitation of the $^{1}S_{0}({\bf 1})$ resonance and the
 ground state of the $^{1}S_{0}({\bf 8_S})$ resonance. \\

\FIGURE[t]{\vspace{10pt}\\
\epsfig{file=mglgl_t.eps,width=0.45\textwidth}
 \caption{Invariant-mass distributions at the threshold region, at the
 LHC, $\sqrt{s}=14$ TeV.
 Results including the all-order Coulomb correction, as well as no
 Coulomb correction are plotted for $m_{\go}=608$ GeV, for
 $\Gamma_{\go}=20$ GeV (dotted), 5.5 GeV (dot-dashed) 1.5 GeV (dashed)
 and 0.5 GeV (solid).\label{fig:thre}}}

Fig.~\ref{fig:thre} is also the invariant-mass distributions at the very
 threshold region, considering a variety of the gluino decay width.
Results for $m_{\go}=608$ GeV, for $\Gamma_{\go}=20$ GeV, 5.5 GeV,
 1.5 GeV and 0.5 GeV are plotted in the dotted, dot-dashed, dashed and
 solid lines, respectively.

According to Fig.~\ref{fig:width}, $\Gamma_{\go}=20$ GeV is the same
 order as the gluinonium binding energy $|E_{B}|\simeq22$ GeV for
 $m_{\go}=608$ GeV.
Thus it corresponds to the region $A$ according to the classification
 defined in Sec.~\ref{sec:dec}.
In this case, the invariant-mass distribution shows no resonance peaks,
 but a gradual slope until rather below the threshold.
This is due to the finite width effect, giving the production
 probability of the events with a off-shell gluino.
The Coulomb corrections act to enhance the production ratio at the
 threshold region.
Next, $\Gamma_{\go}=5.5$ GeV is the same value as used in
 Fig.~\ref{fig:inv} which corresponds to the region $B$.
This case, the distribution shows a broad resonance at $E\simeq-22$ GeV,
 corresponding to the ground $^{1}S_{0}({\bf 1})$ gluinonium.
The resonance for the ground $^{1}S_{0}({\bf 8_S})$ gluinonium at
 $E\simeq-7$ GeV is barely seen.
In the distribution for $\Gamma_{\go}=1.5$ GeV (region $C$), the two
 separated resonances for the ground $^{1}S_{0}({\bf 1})$ and
 $^{1}S_{0}({\bf 8_S})$ gluinoniums can be seen.
Finally, for $\Gamma_{\go}=0.5$ GeV (region $D$), several resonances can
 be found in $M<2m_{\go}$.
Actually, the first peak for the ground $^{1}S_{0}({\bf 1})$ resonance
 is quite sharp.
The second peak at $E\simeq -7$ GeV which consists of the ground
 $^{1}S_{0}({\bf 8_S})$ gluinonium and the second excited
 $^{1}S_{0}({\bf 1})$ gluinonium is also clearly seen.

As mentioned in Sec.~\ref{sec:dec}, for the regions $C$ and $D$, the
 partial decay width of the gluinonium annihilation into gluons can not
 be negligible.
For $m_{\go}=608$ GeV, the two gluon decay-width of the $^{1}S_{0}({\bf
 1})$ gluinonium is about $\Gamma_{gg}[^{1}S_{0}({\bf 1})]\simeq 0.75$
 GeV, see Eq.~(\ref{eq:ggg}) and Fig.~\ref{fig:width}.
Note that the gluinoniums have partial decay width $2\Gamma_{\go}$, due
 to the decay width of the constituent gluinos.
Thus, the branching ratio to the ``hidden'' gluino decay may be
 estimated by $B((\gogo)\to
 gg)=\Gamma_{gg}/(2\Gamma_{\go}+\Gamma_{gg})$.
Using this estimation, about 20\% (40\%) of the resonance events for
 $\Gamma_{\go}=1.5$ GeV (0.5 GeV) in Fig.~\ref{fig:thre} would decay
 into jets.
Therefore, only 80\% (60\%) of the resonance events would leave
 detectable signals of the gluinos at hadron colliders. \\

A correction to the total cross-section is also an important
 consequence of the inclusion of the binding corrections.
Above the threshold, the all-order summation of the Coulomb corrections
 gives the Sommerfeld factor.
It enhances the total cross-section from the Born-level results by 10\%
 to 20\% for $m_{\go}=200$ GeV to 2 TeV, almost independent of the
 gluino decay-width.
This result is rather smaller than that in Ref.~\cite{Kulesza:2009kq},
 due to the NLO correction to the QCD potential which reduces the
 magnitude of the Green's function even for $E\simgt0$.

In addition, there comes a substantial portion to the total cross
 section from the $M<2m_{\go}$ region.
The cross section in this region emerges by (1) the smearing effect due
 to the finite width of the gluino, and (2) the bound-state
 contributions.
The former is proportional to $\Gamma_{\go}$, while the later is almost
 independent of $\Gamma_{\go}$.

\TABLE[b]{\begin{tabular}{|r|r|r|r|r|}
\hline
 $m_{\go}$\hspace{15pt} & $A$: $\Gamma_{\go}=E_{B}$ &
 $B$: $\Gamma_{\go}=E_{B}/2$ & $C$: $\Gamma_{\go}=2\Gamma_{gg}$ &
 $D$: $\Gamma_{\go}=\Gamma_{gg}/2$ \\
\hline \hline
 200 [GeV] &  7.5 [4.5] & 5.0 [1.8] & 4.0 [0.3] & 3.9 [0.1] \\
 400 [GeV] &  7.1 [4.2] & 4.8 [1.7] & 3.8 [0.2] & 3.8 [0.1] \\
 600 [GeV] &  7.2 [4.2] & 5.0 [1.7] & 3.9 [0.2] & 4.2 [0.0] \\
   1 [TeV] &  7.9 [4.6] & 5.5 [1.8] & 4.3 [0.2] & 4.4 [0.0] \\
 1.5 [TeV] &  9.2 [5.3] & 6.3 [2.1] & 5.0 [0.2] & 5.1 [0.0] \\
   2 [TeV] & 10.7 [6.3] & 7.4 [2.5] & 5.9 [0.2] & 5.9 [0.0] \\
\hline
\end{tabular}
\caption{
 Table of a proportion of the cross section in $M<2m_{\go}$ to the total
 cross-section (\%), at the LHC with $\sqrt{s}=14$ TeV.
 The same ratio but without the binding correction are also listed in
 the squark bracket.\label{tab:neg} }}

In Table~\ref{tab:neg}, we calculate the proportion of the
 cross section in $M<2m_{\go}$ to the total cross-section.
We define the total cross-section by integrating the differential
 cross-section over the invariant-mass from
 $M=2m_{\go}+E_{B}[^{1}S_{0}({\bf 1})]-3\Gamma_{\go}$ to $\sqrt{s}$,
 where $E_{B}[^{1}S_{0}({\bf 1})]$ is the binding energy of the
 $^{1}S_{0}({\bf 1})$ gluinonium.
$\Gamma_{\go}=E_{B}[^{1}S_{0}({\bf 1})]$, $E_{B}[^{1}S_{0}({\bf
 1})]/2$, $2\Gamma_{gg}[^{1}S_{0}({\bf 1})]$ and
 $\Gamma_{gg}[^{1}S_{0}({\bf 1})]/2$, for $m_{\go}=200$ GeV to 2 TeV are
 examined.
The four patterns of $\Gamma_{\go}$ represent typical values in the
 regions $A$ to $D$.
In the square brackets are the same ratio but without the binding
 correction, thus these illustrate only the finite-width effect of the
 gluino.

For $\Gamma_{\go}=E_{B}$ (region $A$), the ratio grows from 8\% for
 smaller $m_{\go}$ to 11\% for larger $m_{\go}$, where a more than half
 of the contribution comes from the smearing effect due to the large
 decay width.
The smearing effect decreases with $\Gamma_{\go}$, and is negligible for
 the regions $C$ and $D$.
Then the bound-state effect which tends from 4\% to 6\% dominates the
 ratio.
Basically, the ratio grows with $m_{\go}$.
This is because, for larger $m_{\go}$, the gluon luminosity becomes more
 steep in $M$, thus the cross section in lower invariant-mass region
 is more significant.

\section{ISR corrections}\label{sec:isr}

In this section, we discuss the effects of the ISR and the hard
 correction, which are the remaining major corrections in QCD.
We treat the ISR function to ${\cal O}(\alpha_s)$ in a soft
 approximation, as in Ref.~\cite{Hagiwara:2008df}.
The ISR function defined in Eq.~(\ref{eq:fac}) is written up to NLO as;
\begin{align}
 F^{(c)}_i(z;\mu_F) = \delta{(1-z)}
+ \frac {\alpha_s(\mu_F)}{\pi}\left[
f^{(c)}_{i}\left(z,\frac{\mu_F}{m_{\go}}\right)
+k^{(c)}_{i}
\left(\frac{\mu_F}{m_{\go}}\right)\delta{(1-z)}\right],
\end{align}
 with
\begin{align}
 f^{{(c)}}_i\left(z,\frac{\mu_F}{2m_{\go}}\right) =
 4A_{i}\left[\left(\frac{\ln{(1-z)}}{1-z}\right)_{+}
 - \left(\frac{1}{1-z}\right)_{+}
 \ln{\left(\frac{\mu_F}{2m_{\go}}\right)}\right]
 + D_{c}\left(\frac{1}{1-z}\right)_{+}.
\end{align}
and
\begin{align}
 k^{(c)}_{i}\left(\frac{\mu_F}{m_{\go}}\right) =
 2B_{i}\ln{\left(\frac{\mu_F}{2m_{\go}}\right)}.\label{eq:nonlog}
\end{align}
The constants are given as~\cite{Kulesza:2008jb};
\begin{align}
 & A_{g} = C_A,\quad A_{q} = C_F,\quad
 B_{g} = -\frac{\beta_0}{2},\quad B_{q} = -\frac{3}{2}C_F,\nn\\
 & D_{\bf 1} = 0,\quad D_{\bf 8_S} = -3,\quad
 D_{\bf 8_A} = -3,\quad D_{\bf 27} = -8.
\end{align}
The hard-vertex factor is obtained by matching with the
 full ${\cal O}(\alpha_s)$ correction.
However, to our knowledge, ${\cal O}(\alpha_s)$ correction to
 the gluino-pair production has been calculated numerically and only for
 the color--summed cross-section in Ref.~\cite{Beenakker:1996ch}.
Accordingly, we cannot fixed the color--dependent factors, but only
 the color--averaged one.
We parametrize the hard-vertex factors as
\begin{align}
 K^{(c)}_{i}(\mu_R) = 1 + \frac{\alpha_s(\mu_R)}{\pi}
 h^{(c)}_{i}\left(\frac{\mu_R}{m_{\go}}\right),\label{eq:hard}
\end{align}
with
\begin{align}
 h_{i}\left(\frac{\mu_R}{m_{\go}}\right) = \bar{h}_{i}
 + \beta_0\ln{\left(\frac{\mu_R}{2m_{\go}}\right)}.
\end{align}
Extracting from the graphs in the figure 9 (a) and (b) in
 Ref.~\cite{Beenakker:1996ch} or using Ref.~\cite{Beenakker:1996ed}, we
 find
\begin{align}
 &\bar{h}_{g} = -0.26,\label{eq:h1}\\
 &\bar{h}_{q} = 1.5,\label{eq:h2}
\end{align}
where the later is obtained for $m_{\sq}/m_{\go}=1.4$, and we neglect
 the mass-ratio dependence.
There is also an ambiguity in separating the non-logarithmic part in
 $k^{(c)}_{i}$ and $h^{(c)}_{i}$.
We take a scheme such that the non-logarithmic part in $k^{(c)}_{i}$ is
 zero.\\

Now, we examine the ISR effect to the differential cross section at the
 bound-state region.
We calculate the ratio of $d\sigma/dM$ at $M=2m_{\go}+E_{B}$ including
 the ISR to that without the ISR, $K\equiv\sigma_{\rm ISR}/\sigma_{\rm
No\, ISR}$, where $d\sigma/dM$ with the ISR are evaluated using the
 CTEQ6M PDFs with $\mu=m_{\go}$.

We also examine the ambiguity by the scale choice of $\mu$.
We find the maximum and the minimum of $d\sigma/dM$ at
 $M=2m_{\go}+E_{B}$ during the range from $\mu=m_{\go}$ to $2m_{\go}$,
 and define the uncertainty for the cross sections without and with the
 ISR as
\begin{align}
 \delta_{\{\rm No\,ISR,\,ISR\}}\equiv
 \frac{\sigma_{\rm max}-\sigma_{\rm min}} {\sigma(\mu=m_{\go})}.
\end{align}

\TABLE[b]{\begin{tabular}{|r|c|c|c|}
\hline
 $m_{\go}$ \hspace{10pt} & $K$ & $\delta_{\rm No\,ISR}$ &
 $\delta_{\rm ISR}$ \\
\hline \hline
 200 [GeV] & 1.17 & 0.21 & 0.05 \\
 400 [GeV] & 1.36 & 0.22 & 0.02 \\
 600 [GeV] & 1.52 & 0.23 & 0.01 \\
   1 [TeV] & 1.82 & 0.24 & 0.02 \\
 1.5 [TeV] & 2.25 & 0.25 & 0.03 \\
   2 [TeV] & 2.78 & 0.25 & 0.05 \\
\hline
\end{tabular}
\caption{
 Table of the ratio of the differential cross section $d\sigma/dM$ at
 $M=2m_{\go}+E_{B}$ with the ISR to that without the ISR, as well as the
 scale uncertainties $\delta\sigma/\sigma$.\label{tab:isr}}}

In Table~\ref{tab:isr}, we list the ratio $K$ and the scale
 uncertainties $\delta$ for the cross-section without and with the
 ISR.
In the calculation, the decay width is set to $\Gamma_{\go}=E_{B}/2$,
 however the width dependence is small.
The ratio $K$ is 1.2 for $m_{\go}=200$ GeV and increases with $m_{\go}$
 to 2.8 for $m_{\go}=2$ TeV, due to the terms including the plus
 distribution in $F^{(c)}_{i}$, which take large values when $\hat{s}$
 is close to $M^2$.
The scale uncertainty of the cross section without the ISR is 21-25\%,
 however that with the ISR is remarkably reduced to a few percent.

In the invariant-mass distribution, the size of the ISR correction is
 not actually a constant, but moderately increase with $M$.
As the result, the proportion of the cross section in $M<2m_{\go}$ to
 the total cross section shown in Table.~\ref{tab:neg} is slightly
 reduced. \\

We note several remarks on the ISR.
First, it is pointed out in the top-quark production~\cite{Kiyo:2008bv}
 that the emission of the non-collinear gluon can give the same-order
 correction.
This can be implemented into our analysis by utilizing the NLO
 calculation of the gluinonium production which has not been available
 yet.
Moreover, the color-dependence of the hard-vertex factor is neglected in
 our analysis, which is also resolved by the above calculation or the
 NLO calculation of the gluino-pair production with an explicit color
 projection.
Due to the lack of them, our results have ambiguity in the overall
 magnitude of the ISR correction.
However, we expect that these effects would not considerably deform the
 invariant-mass distribution.

\section{Summary}\label{sec:sum}

In this paper, we study the bound-state corrections in the gluino-pair
 production at hadron colliders.
We perform the all-order summation of the Coulomb corrections using the
 Green's function formalism, and find sizable corrections in the
 gluino-pair invariant-mass distribution as well as in the total
 cross-section.\\

The consequence of the bound-state correction crucially depends on the
 decay width of the gluino.
When $m_{\go}>m_{\sq}$, the gluino decay-width is typically between
 a few and a few hundred GeV; see Fig.~\ref{fig:width}.
In this case, the invariant-mass distribution below and near the mass
 threshold region receive a large enhancement, due to the formation of
 the gluinonium bound-states.
We show that the gluino-pair production cross section below the threshold
 ($M<2m_{\go}$) grows from 4\% to 6\% of the total cross section due to
 the bound-states formation, when the gluino mass increases from 200 GeV
 to 2 TeV.
For the large $\Gamma_{\go}$ case, although the resonance peaks are
 smeared-out, the ratio grows to 8\% to 11\% due to the smearing
 effect.
When $\Gamma_{\go}$ is significantly smaller than the binding energy of
 the gluinoniums, which grows from $\sim 10$ GeV for $m_{\go}\simeq 200$
 GeV to $\sim 50$ GeV for $m_{\go}=2$ TeV (see Fig.~\ref{fig:width}),
 one or two resonance peaks arise in the invariant-mass distribution.
Although the study of the observable signals of the gluino-pair
 production, such as the missing $p_T$ and various kinematical
 observables, is beyond the scope of this report, those events below the
 threshold should be taken into account in the determination of the
 gluino mass.

On the other hand, when $m_{\go}<m_{\sq}$, the gluino decay-width is
 quite tiny; see Fig.~\ref{fig:width}.
In this case, the gluinonium spectra are very sharp and more than two
 resonance peaks are expected (see Fig.~\ref{fig:thre}).
However, the produced gluinonium resonances would decay mainly into
 gluon jets and they may escape detections at hadron collider
 experiments.

\section*{Acknowledgment}
We wish to send special thanks to Y.~Sumino for valuable comments and
discussions.
We also thank organizers of an LHC-focus week meeting at the IPMU
 (Institute for the Physics and Mathematics of the Universe) in March
 2009, where we enjoyed stimulating discussions and received useful
 comments from the participants.
This work is supported in part by the Grant-in-Aid for scientific
research (No.~20340064).


\end{document}